\documentclass[3p,times,preprint]{elsarticle}
\usepackage{graphicx}
\usepackage{siunitx}
\usepackage{wrapfig}
\usepackage{url}
\usepackage{tabularx}
\usepackage{amsmath}
\usepackage{multirow}

\begin{document}

\begin{frontmatter}

\title{Recent application studies of an INTPIX4NA SOIPIX detector-based X-ray camera using an SiTCP-XG 10GbE-based high-speed readout system at KEK facilities}

\author[kek,soken]{Ryutaro NISHIMURA \corref{auth}}
\author[kek,soken]{Noriyuki IGARASHI}
\author[kek,soken]{Daisuke WAKABAYASHI}
\author[kek,soken]{Yuki SHIBAZAKI}
\author[kek]{Yoshio SUZUKI}
\author[kek,soken,tsukuba]{Keiichi HIRANO}
\author[kek,soken]{Hiromi MIKI}
\author[saga]{Akio YONEYAMA}
\author[kek,soken]{Hiroshi SUGIYAMA}
\author[kek]{Kazuyuki HYODO}
\author[kek,soken]{Izumi UMEGAKI}
\author[kek]{Koichiro SHIMOMURA}
\author[kek]{Yasuo ARAI}

\cortext[auth]{ Corresponding author. e-mail: ryutaro.nishimura@kek.jp}

\address[kek]{High Energy Accelerator Research Organization (KEK),\\ Oho 1-1, Tsukuba, Ibaraki, 305-0801, Japan}
\address[soken]{Materials Structure Science Program, Graduate Institute for Advanced Studies, Graduate University for Advanced Studies (SOKENDAI),\\ Oho 1-1, Tsukuba, Ibaraki, 305-0801, Japan}
\address[tsukuba]{Graduate School of Pure and Applied Sciences, University of Tsukuba,\\ Tennodai 1-1-1, Tsukuba, Ibaraki, 305-8571, Japan}
\address[saga]{SAGA Light Source (SAGA-LS),\\ Yayoigaoka 8-7, Tosu, Saga, 841-0005, Japan}

\begin{abstract}
The Silicon-On-Insulator PIXel (SOIPIX) detector is a unique monolithic structure imaging device currently being developed by the SOIPIX group, led by the High Energy Accelerator Research Organization (KEK). 
Our detector team at the KEK Photon Factory (PF) has developed an X-ray camera based on the INTPIX4NA SOIPIX detector. 
This detector provides a sensitive area of 14.1 \si{\times} 8.7 \si{mm^2}, with 425,984 pixels arranged in an 832-column \si{\times} 512-row matrix and a pixel size of 17 \si{\times} 17 \si{\micro\meter^2}, 
and offers high spatial resolution and excellent sensitivity under low-intensity X-ray conditions. 
The readout system used in the X-ray camera is developed at the PF. 
It is equipped with SiTCP-XG, a 10 Gb Ethernet network controller implemented on a field-programmable gate array, enabling high-frame-rate imaging at several hundred hertz. 
We are currently investigating the applicability of this X-ray camera in several experiments at KEK. 
Herein, we report three recent application studies: (1) X-ray zooming microscope optics using two Fresnel zone plates at PF AR-NE1A; (2) phase-contrast X-ray imaging system using a two-crystal X-ray interferometer at PF BL-14C; and (3) nondestructive lithium detection in Li-ion battery electrode materials using muonic X-rays at J-PARC MLF Muon D2. 


\end{abstract}
\end{frontmatter}

\section{Introduction}
\label{sec:intro}
X-ray imaging is a powerful technique for the nondestructive evaluation of material structures. 
Driven by the advancements in X-ray imaging capabilities at synchrotron radiation facilities, we have developed an X-ray camera utilizing the INTPIX4NA detector \cite{intpix4na1, intpix4na2}, which is a member of the SOIPIX detector family \cite{soipix1}.
The X-ray camera has undergone performance evaluations utilizing the High Energy Accelerator Research Organization (KEK) Photon Factory (PF) beamlines \cite{intpix4na2} and is currently being employed in a range of experiments both within and beyond the PF. This study presents three representative application studies: 
(1) X-ray zooming microscope optics employing two Fresnel zone plates (FZPs) \cite{zooming} at PF AR-NE1A. 
(2) Phase-contrast X-ray imaging system utilizing a two-crystal X-ray interferometer \cite{interfero} at PF BL-14C. 
(3) Nondestructive detection of metallic lithium in Li-ion battery electrode materials using muonic X-rays \cite{muon1} at J-PARC MLF Muon D2.

\section{INTPIX4NA Detector-Based X-ray Camera}
\label{sec:i4nacamera}

\subsection{Overview of the INTPIX4NA SOIPIX Detector}

\begin{figure}[t!]
\vspace{.2cm}
\centering
\includegraphics[width=.8\linewidth]{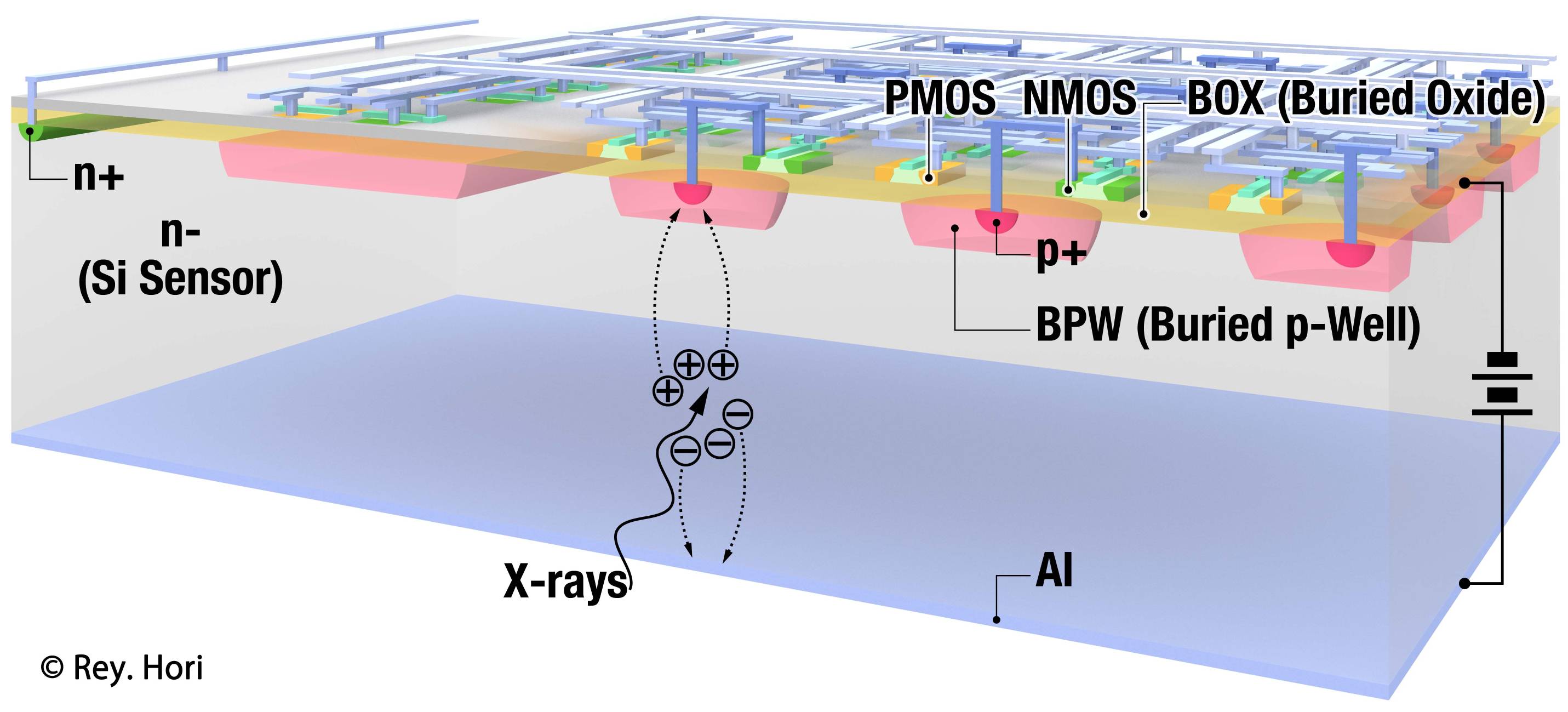}
\caption{Structure of SOIPIX detector.}
\label{fig:soistruct}
\end{figure}

SOIPIX detectors are currently being developed by the SOIPIX collaboration, led by KEK (Tsukuba, Ibaraki, Japan). 
These detectors were fabricated using a 0.2 \si{\micro\meter} CMOS fully depleted silicon-on-insulator (FD-SOI) pixel process developed by Lapis Semiconductor Co., Ltd. \cite{soipix1}. 
The SOIPIX detector comprises a thick, high-resistivity Si substrate for the sensor and a thin Si layer for the CMOS circuits (Figure \ref{fig:soistruct}). 

\begin{figure}[t!]
\vspace{.2cm}
\centering
\includegraphics[width=.8\linewidth]{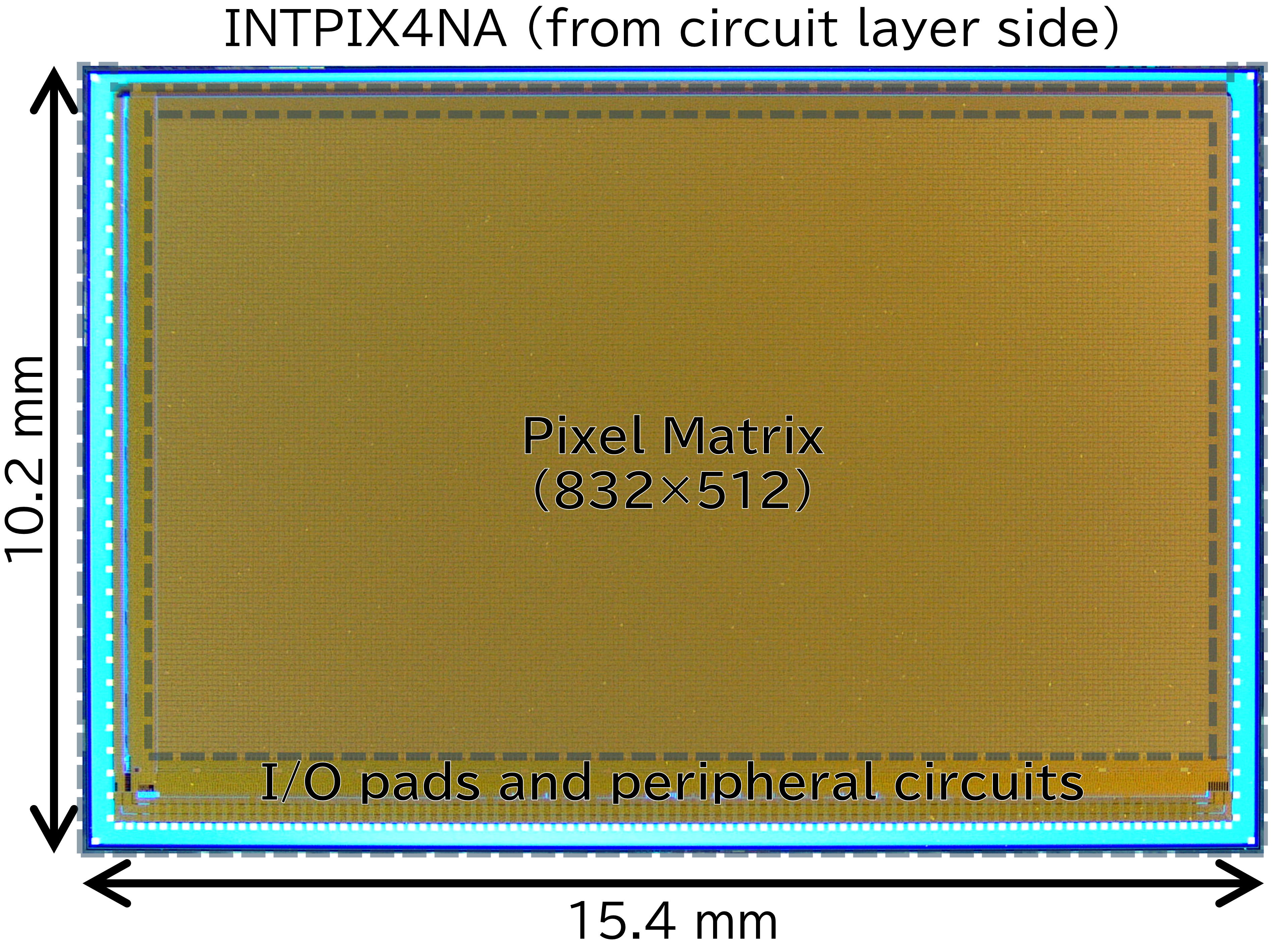}
\caption{Photograph of INTPIX4NA from circuit layer side.}
\label{fig:intpix4na}
\end{figure}

INTPIX4NA \cite{intpix4na1} is a charge-integration-type SOIPIX detector. 
The pixel size is 17 \si{\times} 17 \si{\micro\meter^2}, and the number of pixels is 425,984 (832 columns \si{\times} 512 rows). 
The sensitive area is 14.1 \si{\times} 8.7 \si{mm^2}. 
The design parameters are listed in Table \ref{tab:detector_design}, and a photograph of INTPIX4NA is presented in Figure \ref{fig:intpix4na}. 
Considering its sensitivity and spatial resolution performance in previous studies \cite{intpix4na1, intpix4na2}, INTPIX4NA is suitable for X-ray imaging with 5--20 keV low-intensity X-rays. 

\begin{table}[h]
 \caption{INTPIX4NA design parameters excerpted from prior studies \cite{intpix4na1, intpix4na2}}
 \begin{tabularx}{\linewidth}{XX}
 \hline
 Chip size & 15.4 \si{\times} 10.2 \si{mm^2} \\ 
 Sensitive area & 14.1 \si{\times} 8.7 \si{mm^2} \\ 
 Pixel matrix & 832 columns \si{\times} 512 rows (\SI{425984}{pixels}) \\ 
 Pixel size & 17 \si{\times} 17 \si{\micro\meter^2} \\ 
 Pixel gain & 9.3--10.6 \si{\micro\volt/\elementarycharge} (Actual measurement value) \\ 
 Shutter & Global Shutter \\ 
 Thickness of sensor layer & \SI{300} {\micro\meter} (Typical) \\ 
 Sensor layer wafer & N type Floating Zone wafer \\ 
 Back side Aluminum deposition thickness & \SI{150} {nm} (Typical) \\ 
 Modulation transfer function (MTF) & Over 65 \% at the Nyquist frequency (29.4 cycle/mm) with the 9.6 keV monochromatic X-ray beam. \\ 
 \hline
 \end{tabularx}
 \label{tab:detector_design}
\end{table}

\subsection{Overview of the INTPIX4NA Detector-Based X-ray Camera}

\begin{figure}[t!]
\vspace{.2cm}
\centering
\includegraphics[width=.8\linewidth]{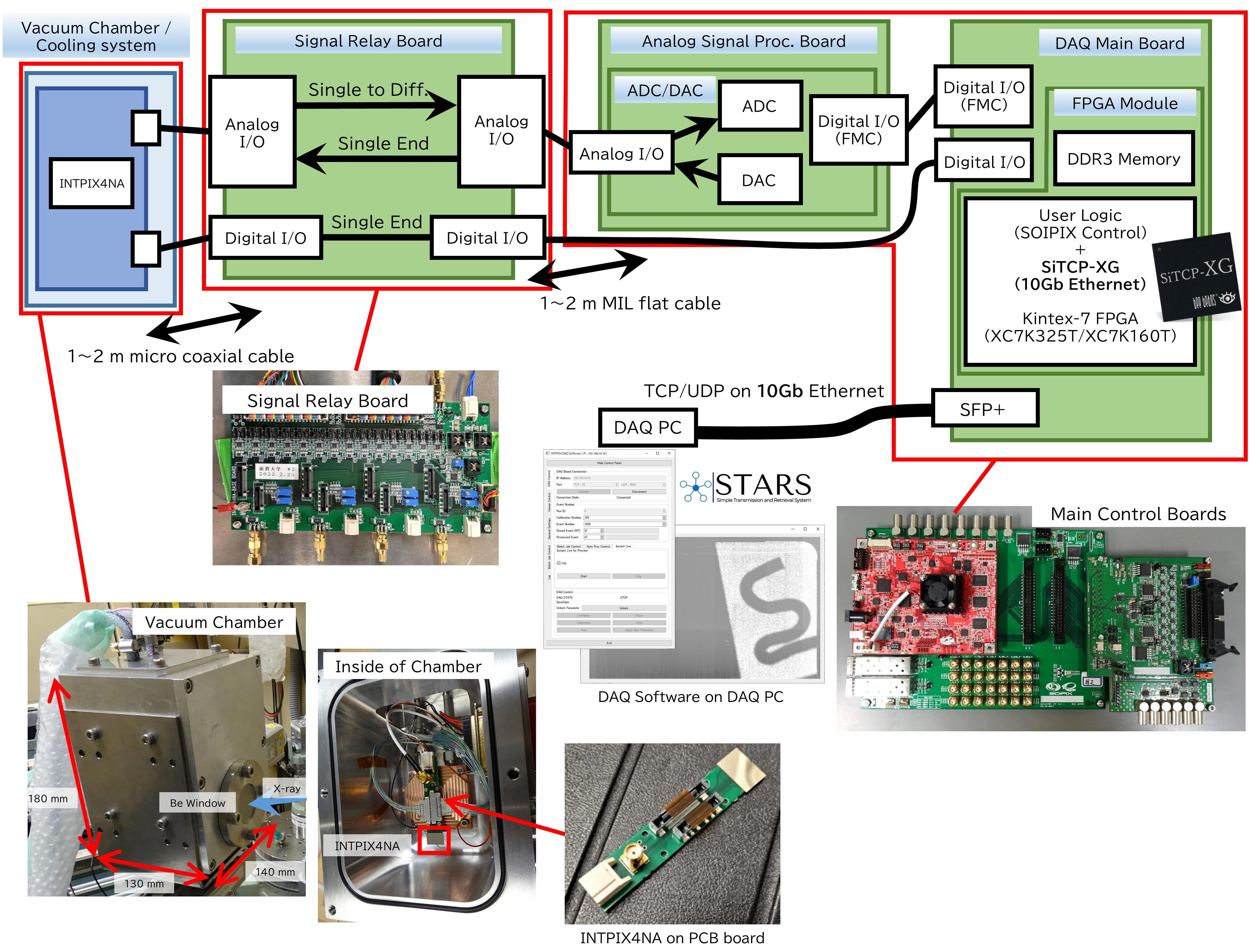}
\caption{Schematics of the INTPIX4NA detector-based X-ray camera. The upper half of this figure shows an overview of the components. The lower half displays photographs of the components and a DAQ software screenshot.}
\label{fig:pfdaqsixschem}
\end{figure}

The INTPIX4NA detector-based X-ray camera consists of a data acquisition (DAQ) board, namely PF-DAQSIX supporting 10 Gbps Ethernet, DAQ software running on a standard PC/AT-compatible computer, a PCB board equipped with an INTPIX4NA detector, and a vacuum chamber equipped with a Peltier element and water cooling system to cool the detector PCB board \cite{intpix4na2}. 
This X-ray camera system supports remote control via a simple transmission and retrieval system (STARS) framework \cite{stars}, enabling coordinated operation with the beamlines at the PF. 
The system schematics, component photographs, and DAQ software screenshot are shown in Figure \ref{fig:pfdaqsixschem}. 

\section{Application Studies and Discussion}

\subsection{Application to the Optics of an X-ray Zooming Microscope Using two Fresnel Zone Plates at PF AR-NE1A}

\begin{figure}[t!]
\vspace{.2cm}
\centering
\includegraphics[width=.8\linewidth]{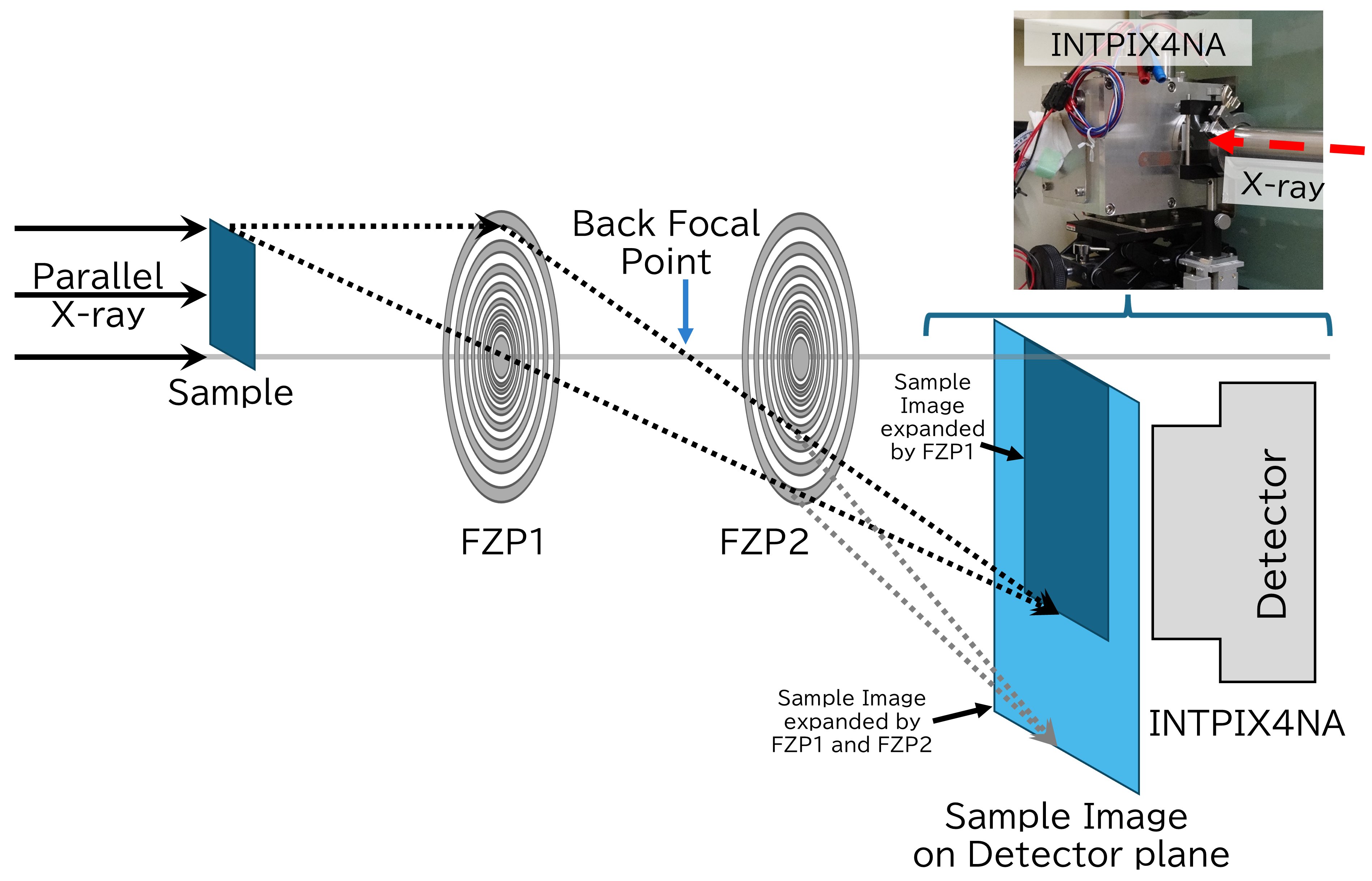}
\caption{Overview of the optics of an X-ray zooming microscope using two FZPs at PF AR-NE1A. In this optical system, the 1-FZP mode, which uses only FZP1, provides low-magnification images (the small, dark-colored sample image shown on the detector plane in the figure). In contrast, the 2-FZPs mode, which combines FZP1 and FZP2, enables high-magnification images (the large, bright-colored sample image shown on the detector plane in the figure).}
\label{fig:zooming}
\end{figure}

An X-ray zooming microscope using two FZPs \cite{zooming} (Figure \ref{fig:zooming}) forms the optics for X-ray microscopy, enabling zooming by changing only the positions of the two FZPs while maintaining a fixed camera length. 
Because high magnification can be achieved with a short camera length, these optics are broadly applicable and initially installed at PF AR NE1A \cite{ar-ne1a}. 
Using this optical system, we conducted experiments, including computed laminography \cite{laminography} and schlieren phase-contrast microscopy \cite{schlieren1}, employing the INTPIX4NA based camera as an imaging device. 
Previous studies \cite{intpix4na2} have demonstrated that, even under the low-intensity and low-contrast conditions such as those arising from significant FZP absorption in the X-ray zooming microscope, the INTPIX4NA-based camera achieves excellent resolution performance consistent with that predicted from its specifications. 
Here, we present two examples of the experiments. 
(1) Computed laminography of a high-pressure sample in diamond anvil cell (DAC) \cite{laminography_dac}. 
(2) Schlieren phase-contrast microscopy of fiber texture in Japanese traditional paper ``Washi.''

\subsubsection{Computed Laminography of High-Pressure Sample in a Diamond Anvil Cell}

A diamond anvil cell (DAC) is a useful tool for reproducing high-pressure and high-temperature conditions corresponding to those in the deep interior of the Earth \cite{laminography_dac}. 
Performing three-dimensional imaging measurements using an X-ray computed laminography technique is a powerful method for acquiring physical information, such as pressure-induced changes in the shape and texture of high-pressure samples in the DAC.
At PF AR-NE1A, high-magnification X-ray computed laminography was conducted for DAC samples using an X-ray zooming microscope. 

\begin{figure}[t!]
\vspace{.2cm}
\centering
\includegraphics[width=.8\linewidth]{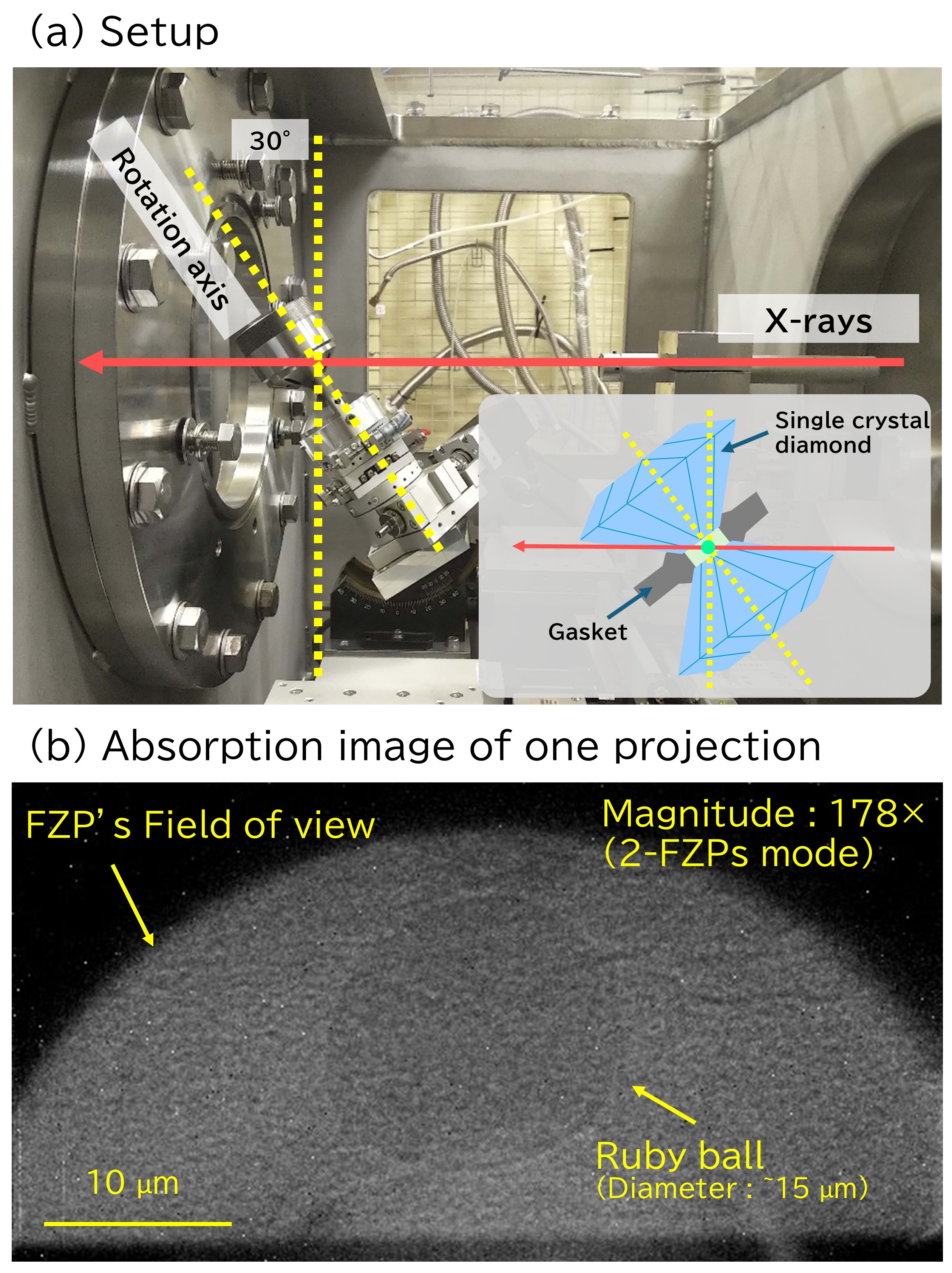}
\caption{Computed laminography of a high-pressure sample in the DAC (E=9.6 keV). (a) Setup of DAC for laminography measurement. (b) One shot of the small ruby ball sample in the DAC, from laminography dataset.}
\label{fig:zooming_dac_result}
\end{figure}

In this section, we present an example of an X-ray computed laminography measurement performed under the conditions of a full-rotation scan (\ang{0.5} steps, 721 data sets) on a small ruby ball sample within a panoramic DAC. 
This measurement was performed with 2-FZPs mode (magnitude: 178\si{\times}), and the X-ray energy was 9.6 keV. 
The DAC featured two symmetric unobstructed radial openings of \ang{150} each in the horizontal direction, providing a total opening of \ang{300}. 
The diamond culet size and applied pressure were 300 {\textmu}m and 10 GPa, respectively. 
The gasket material was W-Re (thickness: 40 {\textmu}m; hole diameter: 120 {\textmu}m). 
The sample was a ruby ball (diameter: approximately 15 {\textmu}m) with a cBN-powder pressure medium. 
Figure \ref{fig:zooming_dac_result} (a) shows the setup of the DAC for laminography measurements on the upper side of the X-ray zooming microscope. 
Figure \ref{fig:zooming_dac_result} (b) shows a shot of a small ruby ball sample in the DAC, extracted from the laminography dataset. 
Although measurements were performed at low X-ray intensities owing to the use of two FZP plates under these conditions, 
the INTPIX4NA detector-based X-ray camera captured the subtle differences in contrast. 

\subsubsection{Schlieren Phase-Contrast Microscopy of Fiber Texture in Japanese Traditional Paper ``Washi''}

\begin{figure}[t!]
\vspace{.2cm}
\centering
\includegraphics[height=.95\textheight, keepaspectratio]{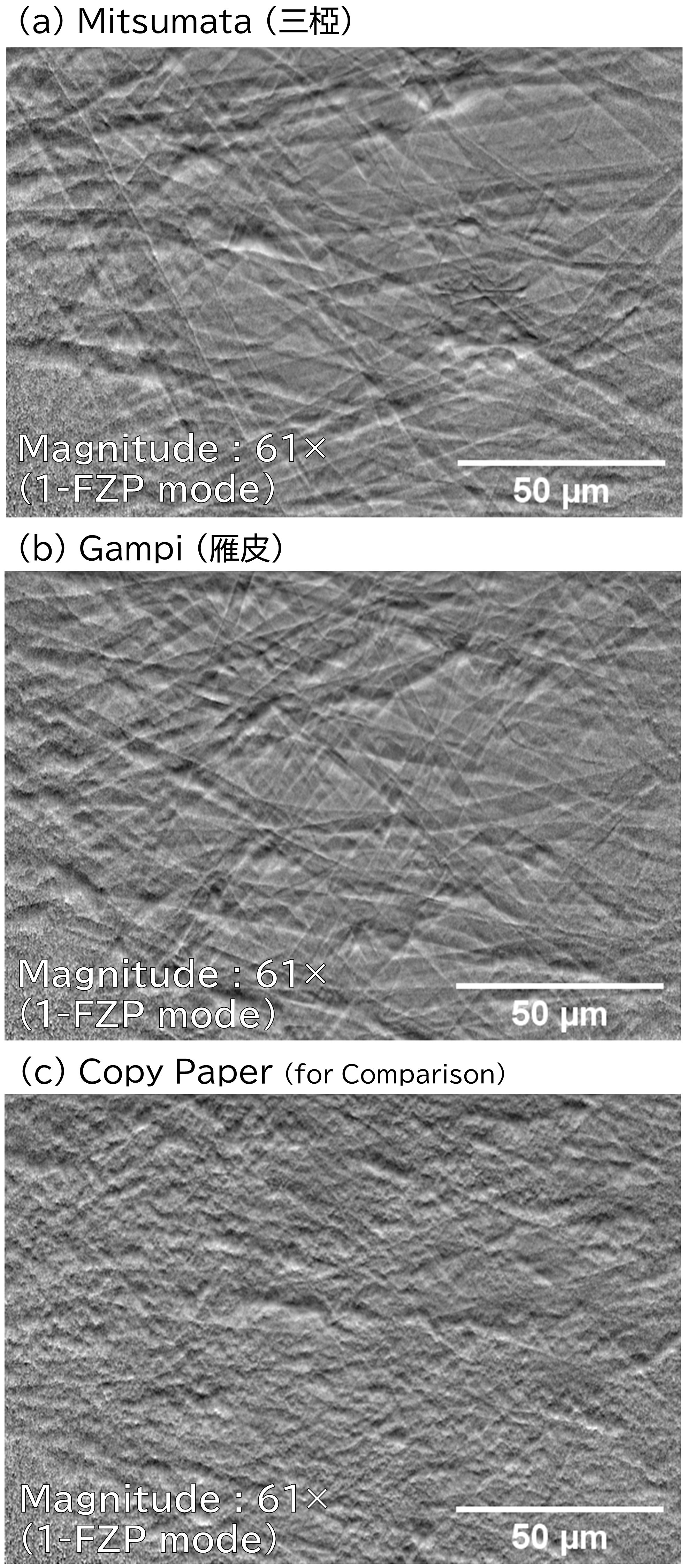}
\caption{Phase-recovered images of the fiber structure of Japanese traditional paper ``Washi'' obtained by schlieren phase-contrast microscopy with 9.6keV monochromatic X-ray. (a) Mitsumata. (b) Gampi. (c) Copy paper for comparison.}
\label{fig:zooming_jpnpaper_result}
\end{figure}

In the X-ray region, the phase contrast generated by the real part of the complex refractive index is generally significantly higher than that generated by the imaginary part \cite{momose1}. 
Owing to the detection of phase shifts, several phase contrast methods have been developed to convert them into contrast, and the schlieren method is one such technique. 
In the X-ray zooming microscope, the schlieren phase contrast can be generated by excluding one side of the diffracted waves by placing a single-blade beam stop (knife-edge filter) at the back focal plane of the objective lens (FZP1 in Figure \ref{fig:zooming}) \cite{schlieren1}. 

In this section, we present the results of schlieren phase-contrast imaging measurements conducted on two types of traditional Japanese paper, known as Washi. 
These samples are composed of light elements; therefore, achieving contrast through absorption imaging is challenging. 
Consequently, phase imaging is considered suitable for such measurements. 
This measurement was performed using 1-FZP mode (magnitude: 61\si{\times}), and the X-ray energy was 9.6 keV. 
Figure \ref{fig:zooming_jpnpaper_result} shows the phase-recovered images of the fiber structure of Japanese traditional paper ``Washi'' obtained by this schlieren phase-contrast microscopy measurement. 
Phase recovery for these results was performed using a phase-retrieval algorithm applied to a single-shot image \cite{schlieren2}.
Figure \ref{fig:zooming_jpnpaper_result} (a) shows the result of the sample known as ``Mitsumata.'' 
Figure \ref{fig:zooming_jpnpaper_result} (b) shows the result of the sample known as ``Gampi.'' 
Figure \ref{fig:zooming_jpnpaper_result} (c) shows the results of copy paper for comparison. 
In such phase recovery processing, high-sensitivity and high-linearity detection of the phase contrast by the detector is crucial. 
The results of the phase recovery processing, as illustrated in Figure \ref{fig:zooming_jpnpaper_result}, reveal variations in the fiber structure among different paper types. 
This result demonstrates that the INTPIX4NA detector-based X-ray camera exhibits the required performance for the phase recovery processing algorithm. 

\subsection{Application to a Phase-Contrast X-ray Imaging System Using a two-Crystal X-ray Interferometer at PF BL-14C}

\begin{figure}[t!]
\vspace{.2cm}
\centering
\includegraphics[width=.8\linewidth]{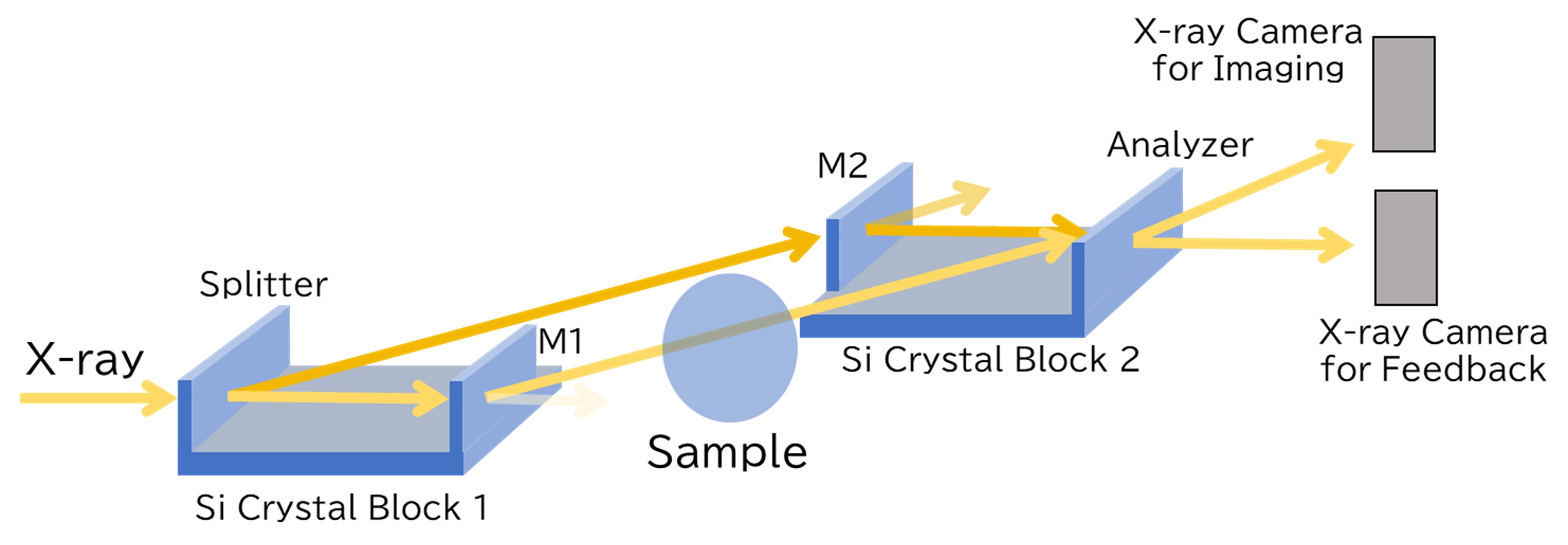}
\caption{Setup of two-crystal X-ray interferometer optics.}
\label{fig:x-ray_interferometer}
\end{figure}

The crystal-based X-ray interferometer detects phase shifts using crystal-based X-ray interferometry, a phase contrast technique \cite{interfero}. 
It offers high sensitivity compared to other methods by directly detecting sample phase shifts. 
Figure \ref{fig:x-ray_interferometer} shows a schematic view of the screw symmetric two-crystal X-ray interferometer. 
The incident X-ray divides into two beams at the splitter crystal (the left edge of Si crystal block 1). 
One beam is reflected by the M1 mirror crystal (the right edge of Si crystal block 1), 
while the other is reflected by the M2 mirror crystal (the left edge of Si crystal block 2). 
The beams superimpose on the analyzer crystal (the right edge of Si crystal block 2) creating two interference beams, 
one for phase-contrast imaging and one for the feedback system. 

\begin{figure}[t!]
\vspace{.2cm}
\centering
\includegraphics[width=.8\linewidth]{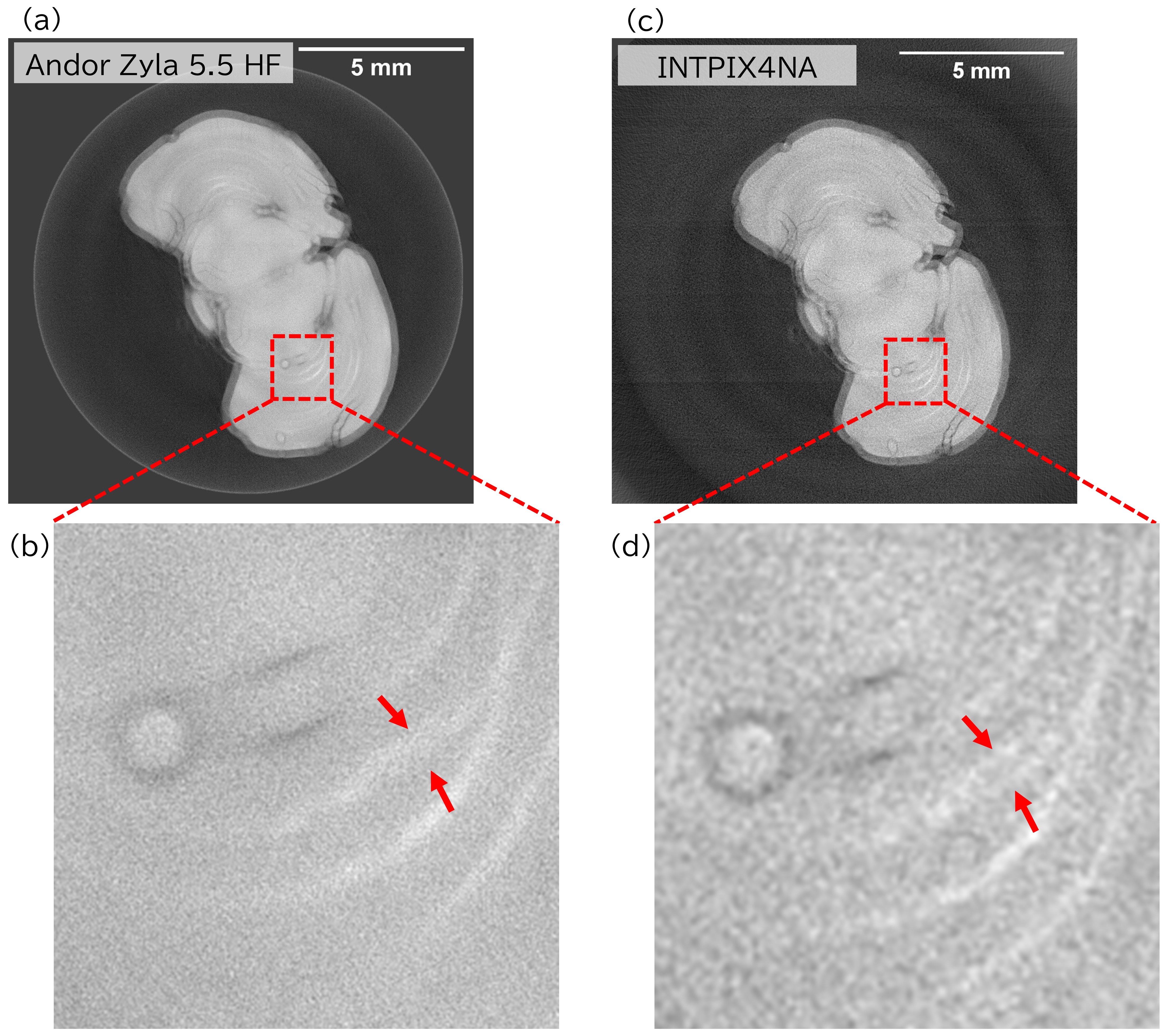}
\caption{Phase-contrast CT slice of a mouse brain (E = 17.8 keV). (a) CT slice reconstructed from a dataset acquired using an Andor Zyla 5.5 HF X-ray sCMOS camera. (b) Expanded image of the area indicated by the dotted line in (a). (c) CT slice reconstructed from a dataset acquired using an INTPIX4NA detector-based X-ray camera. (d) Expanded image of the area indicated by the dotted line in (c). The arrows shown in (b) and (d) indicate the same point of tissue boundaries that are closely spaced.}
\label{fig:x-ray_interferometer_resut}
\end{figure}

In this section, we present the measurement results of phase-contrast computed tomography (CT) slices of a mouse brain (E=17.8keV) using a two-crystal X-ray interferometer at PF BL-14C \cite{bl-14c}. 
Phase-contrast images were acquired using a conventional fiber-coupled X-ray camera (Andor Zyla 5.5 HF, equipped with a 6.5 \si{\micro\meter} square pixel, 2560 \si{\times} 2160 pixel array X-ray sCMOS detector), and an INTPIX4NA detector-based X-ray camera, and the reconstructed CT slices were compared. 
This measurement was performed with a 17.8 keV monochromatic X-ray. 
Figure \ref{fig:x-ray_interferometer_resut} shows the results for phase-contrast CT slices of a mouse brain. 
Figure \ref{fig:x-ray_interferometer_resut} (a) shows a CT slice reconstructed from a dataset acquired using an Andor Zyla 5.5 HF X-ray sCMOS camera. 
Figure \ref{fig:x-ray_interferometer_resut} (b) shows an expanded image of the area indicated by the dotted line in figure (a). 
Figure \ref{fig:x-ray_interferometer_resut} (c) shows a CT slice reconstructed from a dataset acquired using an INTPIX4NA detector-based X-ray camera. 
Figure \ref{fig:x-ray_interferometer_resut} (d) shows an expanded image of the area indicated by the dotted line in figure (c). 
The arrows shown in figures (b) and (d) indicate the same points where the tissue boundaries are closely spaced. 
In figure (b), the tissue boundaries at the indicated point are not clearly resolved and appear as a single blurred line. 
In contrast, in figure (d), these tissue boundaries were resolved as two distinct boundaries. 
A comparison of these slices, particularly the expanded images in (b) and (d), confirms that the INTPIX4NA detector-based X-ray camera captures tissue boundaries more clearly than the Andor Zyla 5.5 HF. 

\subsection{Application to Nondestructive Detection of Metallic Lithium in Li-ion Battery Electrode Materials Using Muonic X-rays at J-PARC MLF Muon D2}

\begin{figure}[t!]
\vspace{.2cm}
\centering
\includegraphics[width=.8\linewidth]{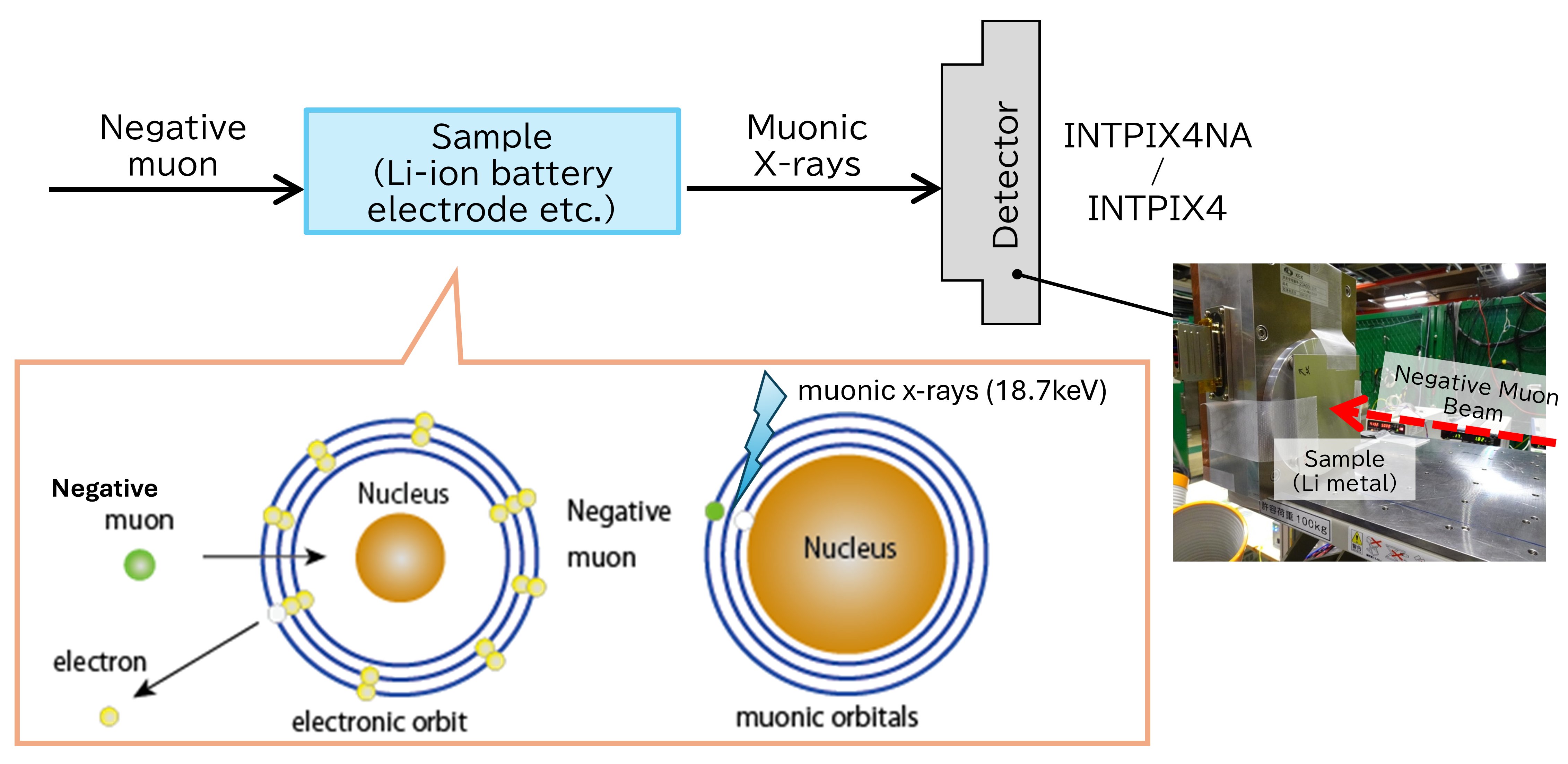}
\caption{Schematic of nondestructive detection of metallic lithium in Li-ion battery electrode materials using muonic X-rays.}
\label{fig:li_muon_setup}
\end{figure}

Lithium-ion batteries are increasingly being applied across a wide range of fields such as portable electronic devices and electric vehicles owing to their electrical capacity and durability. 
In principle, the lithium in a lithium-ion battery always exists in the lithium-ion ($\rm{Li^+}$) state. 
However, in an overcharged state or at temperatures below 5 \si{\degreeCelsius}, $\rm{Li^+}$ ions are irreversibly transformed into metallic lithium ions. 
Metallic lithium usually segregates at the anode and induces severe problems in lithium-ion batteries, such as capacity fading and safety issues. 
Therefore, understanding the state of metallic lithium deposition within lithium-ion batteries is crucial for improving battery performance and establishing health assessment technologies. 
Nondestructive detection of lithium in lithium-ion battery electrode materials using muonic X-rays \cite{muon1} is an advanced technique for understanding the state of metallic lithium deposition in lithium-ion batteries. 
It utilizes the difference in the atomic Coulomb capture ratio of negative muons between lithium metal and lithium ions to measure the distribution of metallic lithium by capturing muon X-rays emitted from metallic lithium within lithium-ion batteries (figure \ref{fig:li_muon_setup}). 
At the J-PARC Materials and Life Science Experimental Facility (MLF) Muon D2 beamline, efforts are underway to integrate multiple two-dimensional semiconductor detectors, including an X-ray camera based on the INTPIX4NA detector. 

\begin{figure}[t!]
\vspace{.2cm}
\centering
\includegraphics[width=.8\linewidth]{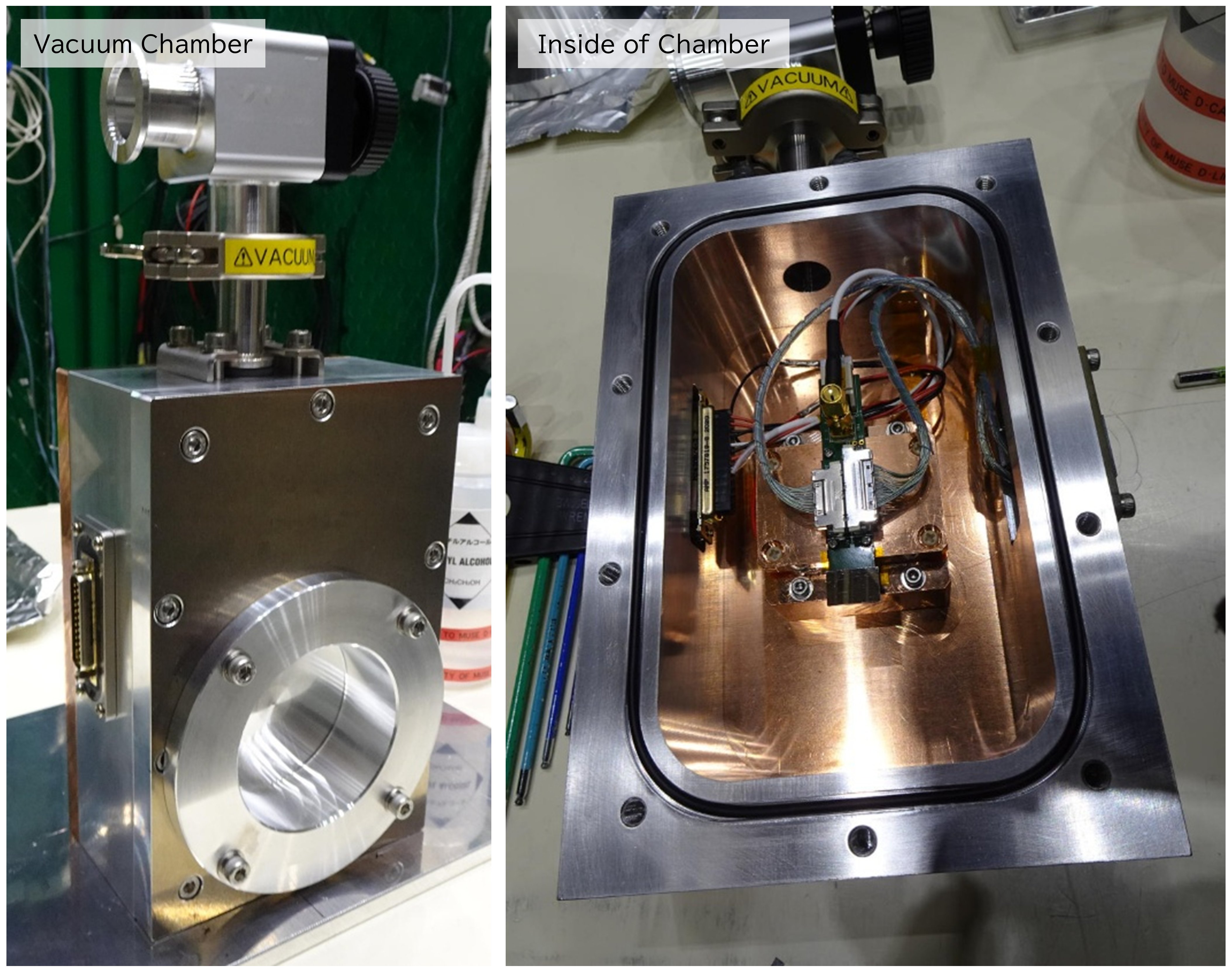}
\caption{Photograph of the customized vacuum chamber for muonic X-ray measurement.}
\label{fig:li_muon_chamber}
\end{figure}

For this experiment, a vacuum chamber with a cooling system was developed based on the one developed by the PF (shown at the bottom left of Figure \ref{fig:pfdaqsixschem}). 
This new vacuum chamber incorporates partial simplification and miniaturization, along with customization to meet the cooling requirements and sample setup. 
Figure \ref{fig:li_muon_chamber} shows a photograph of the customized vacuum chamber used for muonic X-ray measurements. 

\begin{figure}[t!]
\vspace{.2cm}
\centering
\includegraphics[width=.8\linewidth]{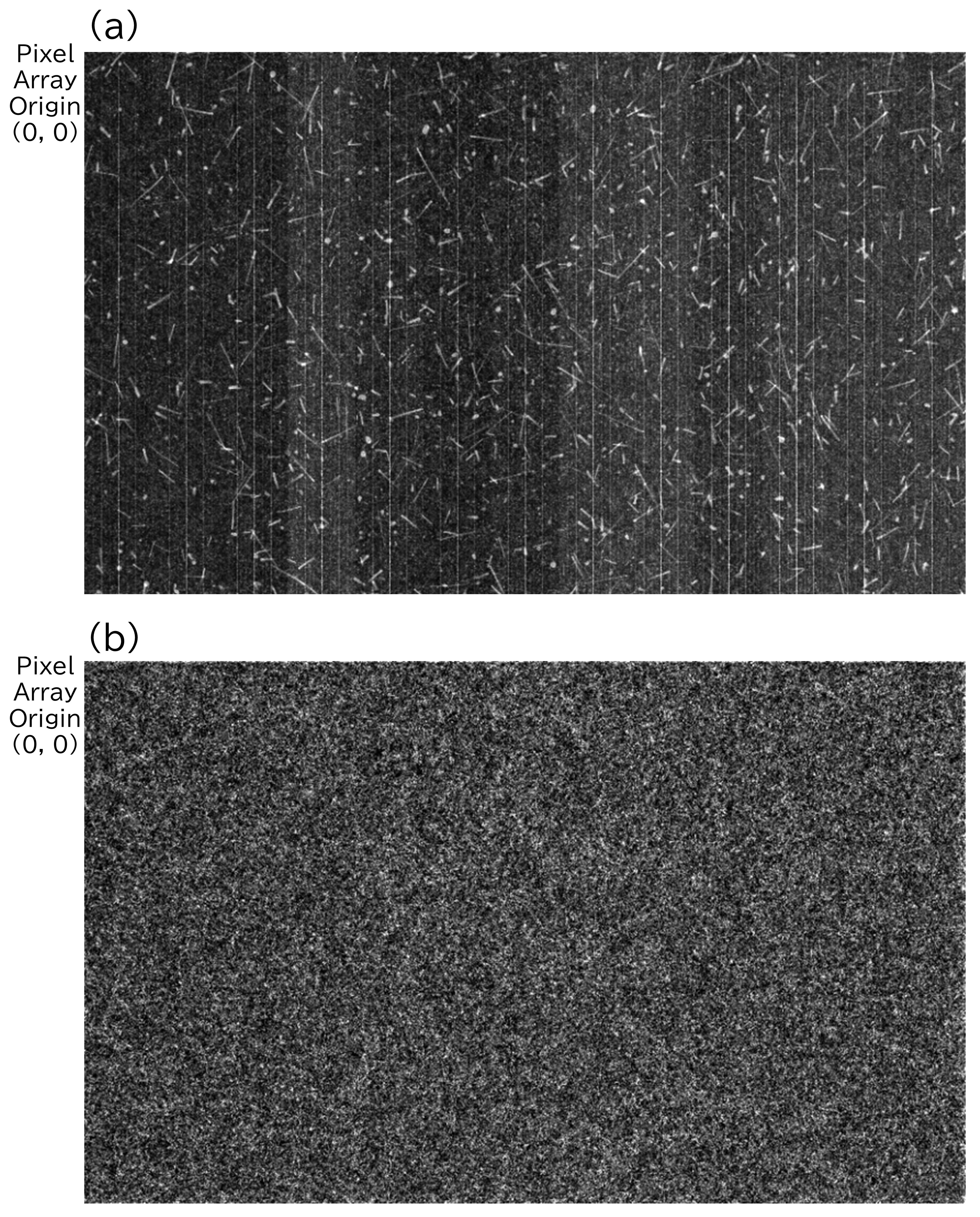}
\caption{Initial result of muonic X-ray detection measurement with lithium metal bulk sample (Momentum : 23.5 MeV/c). (a) 0.3 \si{\micro s} \si{\times} 60000 frames stacked image including all detected particles. (b) 0.3 \si{\micro s} \si{\times} 60000 frames stacked image excluded non-X-ray signals.}
\label{fig:li_muon_result}
\end{figure}

This application is currently being investigated and developed. 
Figure \ref{fig:li_muon_result} shows the initial results of muonic X-ray detection measurement with the bulk lithium metal sample (negative muon momentum of 23.5MeV/c). 
Figure \ref{fig:li_muon_result} (a) shows a 0.3 \si{\micro s} \si{\times} 60000 frames stacked image, including all detected particles. 
Figure \ref{fig:li_muon_result} (b) shows a 0.3 \si{\micro s} \si{\times} 60000 frames stacked image, excluding the non-X-ray signals from each frame. 
Because this detector is charge-integrating, it simultaneously collects charges from various incident charged particles along with muon X-rays. 
Integrating these without distinguishing between them yields the results shown in Figure \ref{fig:li_muon_result} (a). 
However, by extracting and integrating only the X-ray-like signal using the collected charge and spread width, the results shown in Figure \ref{fig:li_muon_result} (b) are obtained. 
This result suggests that muon X-rays can be separated using relatively simple processing. 
Although the peak positions in the spectra currently obtained from these data are consistent with expectations, the spectral shapes are broader and less distinct than anticipated. This is believed to result from insufficient gain calibration between pixels and the influence of electrical noise. Therefore, efforts to address this issue are ongoing. 

\section{Conclusions}

In this study, we present the recent application results obtained using an INTPIX4NA SOIPIX detector–based X-ray camera equipped with a SiTCP XG 10 GbE high speed readout system. 
The detector offers high spatial resolution and exhibits high sensitivity to low-intensity X-rays. Efforts are underway to extend its application both within and beyond the PF. 

First, the X-ray zooming microscope optics, equipped with two FZPs at PF ARNE1A, successfully captured subtle contrast variations in high-pressure samples within a diamond anvil cell (DAC). 
This setup facilitated phase-contrast imaging of samples composed of light elements, such as traditional Japanese paper.
These results confirm that the high resolution, high sensitivity, and linear response characteristics of the detector are conducive to performing measurements in low-intensity environments within X-ray zooming microscope optics.

Second, in phase-contrast X-ray imaging utilizing the two-crystal X-ray interferometer at PF BL14C, the camera facilitated enhanced visualization of tissue boundaries in mouse brain CT slices, thereby outperforming conventional fiber-coupled sCMOS systems. 
This result demonstrates the advantage of the direct conversion SOIPIX architecture for high precision phase contrast CT. 

Finally, preliminary experiments conducted at the JPARC MLF Muon D2 facility for muonic X-ray detection demonstrated that X-ray-like signals originating from metallic lithium can be extracted from charge-integrated images through straightforward processing techniques. 
This finding underscores the significant potential of the detector for the nondestructive evaluation of Li-ion battery materials. 

These results collectively demonstrate that the INTPIX4NA detector-based X-ray camera is a versatile and powerful imaging tool capable of supporting diverse and advanced experimental techniques at synchrotron and muon facilities. 
Ongoing developments will further expand its applicability to high-sensitivity, high-resolution X-ray imaging and related scientific fields. 

\section{Acknowledgements}
Part of this study was conducted at KEK Photon Factory beamlines under the approval of the Photon Factory Program Advisory Committee (Proposal No. 2021PF-S001, 2024PF-G003, 2024PF-G006, 2025PF-G004 and 2025PF-G016). 
Part of this study was conducted at the Materials and Life Science Experimental Facility of the J-PARC under a user program (Proposal No. 2022MS01).
This work was supported by JSPS KAKENHI Grant Number 21K14174, 22H03875, 23K25129, 24K00740 and 25K22028. 

\section*{CRediT authorship contribution statement}
\textbf{Ryutaro NISHIMURA:} Conceptualization, Data curation, Formal analysis, Funding acquisition, Investigation, Methodology, Project administration, Resources, Software, Visualization, Writing - Original Draft, Writing - Review \& Editing. 
\textbf{Noriyuki IGARASHI:} Funding acquisition, Project administration, Supervision, Writing - Review \& Editing. 
\textbf{Daisuke WAKABAYASHI:} Data curation, Funding acquisition, Investigation, Methodology, Resources, Validation, Writing - Review \& Editing. 
\textbf{Yuki SHIBAZAKI:} Data curation, Funding acquisition, Investigation, Methodology, Resources, Validation, Visualization, Writing - Review \& Editing. 
\textbf{Yoshio SUZUKI:} Data curation, Investigation, Methodology, Resources, Validation, Writing - Review \& Editing. 
\textbf{Keiichi HIRANO:} Project administration, Resources, Supervision, Validation, Writing - Review \& Editing. 
\textbf{Hiromi MIKI:} Data curation, Investigation, Resources, Writing - Review \& Editing. 
\textbf{Akio YONEYAMA:} Data curation, Investigation, Resources, Visualization, Writing - Review \& Editing. 
\textbf{Hiroshi SUGIYAMA:} Investigation, Resources, Writing - Review \& Editing. 
\textbf{Kazuyuki HYODO:} Resources, Writing - Review \& Editing. 
\textbf{Izumi UMEGAKI:} Data curation, Funding acquisition, Investigation, Methodology, Resources, Writing - Review \& Editing. 
\textbf{Koichiro SHIMOMURA:} Project administration, Supervision, Writing - Review \& Editing. 
\textbf{Yasuo ARAI:} Resources, Writing - Review \& Editing. 


{}

\end{document}